# Numerical Solutions of Heat Diffusion Equation Over One Dimensional Rod Region

Mehran Makhtoumi
Dept. of Aerospace Engineering
Science and Research Branch, IAU
Tehran, Iran
mehran.makhtoumi@srbiau.ac.ir

*Abstract*— **Adaptive methods for derivation of analytical and numerical solutions of heat diffusion in one dimensional thin rod have investigated. Comperhensive comparsion analysis based on the homotopy perturbation method (HPM) and finite difference method (FDM) have been applied to the rod PDE system. The results show that performing HPM will eventuate more precision and satisfactory approximations at reasonable time than those obtained from FDM when compared to exact solution results. Also since solutions are originated from the problems in HPM thus it is convenient to express them with different functions which conclude that homotopy perturbation is a powerful numerical technique for solving partial differential equations.**

*Keywords- Homotopy perturbation, Finite difference, Heat diffusion.*

## I. INTRODUCTION

In a thin rod with non uniform temperature, thermal energy is transferred from regions of higher temperature to regions of lower temperature. Heat diffusion equation is a parabolic partial differential equation which describes the heat distribution in a given region and provides the basic tool for heat conduction analysis. The solution to this equation provides knowledge of the temperature distribution which may then be used with Fourier's law to determine the heat flux [1-3]. Analytical and numerical methods have gained the interest of researchers for finding approximate solutions to PDEs and this interest is driven by demand of applications both in industry and sciences which leads to investigate analytical and numerical methods for solving initial and boundary value problems [4].

From among of many advanced numerical methods applied to calculation most often finite difference method (FDM), finite element method (FEM), boundary element method (BEM) and finite volume method (FVM) are used. As for instance, Skrzypczak et al. (2008) implemented FEM and DGM for transient heat transfer problem with Newton boundary condition in square testing domain. The results obtained by them shows that discontinuous classical Galerkin finite element method applied to higher order approximation allows to obtain accurate results even on coarse grids and appropriate combination of spatial adaptation of the mesh with increasing of polynomial order of basis functions leads to accurate solutions with efficient memory usage and computational cost [5-7]. One of the widely applied techniques are perturbation methods. In 1999, Ji Huan has proposed a new perturbation technique coupled with the homotopy technique, which is called the homotopy perturbation method (HPM).

Hemeda (2012) studied homotopy perturbation method for solving systems of nonlinear coupled equations. He used HPM to solve some systems of partial differential equations viz. the systems of coupled Burgers' equations in one and two dimensions and the system of Laplace's equation. The results obtained in this study confirmed the power, simplicity and efficiency of HPM compared with the other numerical methods. The author also concluded that the HPM is a suitable method for solving any partial differential equation or systems of partial differential equations as well [11]. Recently, Cheniguel (2014) applied the homotopy perturbation method for solving different one dimensional heat conduction problems with dirichlet and neumann boundary conditions. Cheniguel concluded that the problems solved by using HPM give satisfactory results in comparison to those recently obtained by Yambangwai et al. (2013) which used utilized deferred correction method to increase the order of spatial accuracy of the crank-nikolson scheme for the numerical solution of the one-dimensional heat equation [4,10]. In 2015, Xiao Jun et al. proposed a local fractional homotopy perturbation method which was the extended form of the classical homotopy perturbation method. They investigated the effectiveness of local fractional homotopy perturbation method and its convergence by applying two different examples. The results of this study proved that the new method has high accuracy and can be applied to other partial differential equations [13]. Also, Yasir Khan and Qingbiao (2011) proposed a combined form of the Laplace transform method with the homotopy perturbation method to solve nonlinear equations. They called the new method as homotopy perturbation transform method (HPTM). In this study, the proposed scheme found the solutions without any discretization or restrictive assumptions and avoided the round off errors. In addition the fact that the proposed technique solved nonlinear problems without using Adomian's polynomials could be considered as a clear advantage of this algorithm over the decomposition method [14]. Changbum Chun et al. (2010) examined homotopy perturbation technique for solving two point boundary value problems with comparison of other methods. They implemented the homotopy perturbation method for solving the linear and nonlinear two point boundary value problems to compare the performance of the homotopy perturbation method with extended Adomian decomposition method and shooting method. In this study using HPM yielded relatively more accurate results with rapid convergence than other methods [15]. Most recently, Mritunjay et al. (2017) developed solution of one dimensional space and time fractional advection dispersion equation by homotopy perturbation method. They considered various forms of dispersion and velocity profiles such as space dependent and both space–time dependent throughout the study. The results of this study conclude that the main advantage of using HPM is that it does not require much information about the boundary of the aquifer and also, the results obtained by HPM can be made more accurate, depending upon the number of terms considered in the approximate analytical solution [12].

In this study, theoretical framework of the heat diffusion in one dimensional thin rod is introduced and then the basic idea of analytical solution using Fourier series with boundary and initial condition assumption is presented. The methods of numerical solutions used in this study are the homotopy perturbation and the finite difference methods, well addressed in references [8-9]. The HPM has attracted the attention of researchers in recent years and here we will show that the main advantage of this method is the fact that it produces the approximate solutions quite fast and the results obtained by HPM are more reliable than those obtained by FDM.

## II. THEORETICAL FRAMEWORK

### A. Analytical Solution Using Fourier Series

In this section solutions to the second order PDE of one dimensional heat equation for the rod shown in Fig (1) is presented. The presented heat equation simulates thermal flux in a thin rod which is insulated except at the two ends. Spatial and time variable functions are solutions of this equation. Since there is only one spatial dimension, the differential equation is refered to as "one-dimensional" in description. The general form of partial differential diffusion equation can be expressed as:

$$\frac{\partial U}{\partial t} = k \nabla^2 U \qquad (1)$$

Here, $k$ is the thermal diffusivity and U the temperature. Physically, the one-dimensional heat conduction equation is given by

$$\frac{\partial U}{\partial t} = k \frac{\partial^2 U}{\partial x^2} \qquad (2)$$

Seperation of variabels which is written in Eq. (3) is used to solve Eq. (2); thus, Eq. (4) is obtained:

$$U(x,t) = X(x)\, T(t) \qquad (3)$$

$$\frac{1}{kT}\frac{dT}{dt} = \frac{1}{X}\frac{d^2X}{dx^2} = -\frac{1}{\lambda^2} \qquad (4)$$

In Eq. (4) each side is equal to a constant. Since the solution must be finite at all times, exponential solution in T is considered to be a negative constant thus the T and X solution are written

$$T(t) = A\, e^{-kt/\lambda^2} \qquad (5)$$

$$X(x) = B \cos(\tfrac{x}{\lambda}) + C \sin(\tfrac{x}{\lambda}) \qquad (6)$$

Substituting Eq. (5) and (6) in Eq. (3) the general solution is given

$$U(x,t) = e^{-kt/\lambda^2}\left[D \cos(\tfrac{x}{\lambda}) + E \sin(\tfrac{x}{\lambda})\right] \qquad (7)$$

### B. Boundary and Initial Conditions

To benefit the heat equation, boundary conditions which are the temperature of rod at ends are considered as follows

$$U(0,t) = U(L,t) = 0 \qquad (8)$$

Applying Eq. (8) in Eq. (7) the general solution becomes:

$$U(x,t) = \sum_{n=1}^{\infty} c_n \sin\left(\tfrac{n\pi x}{L}\right) e^{-k\left(\tfrac{n\pi}{L}\right)^2 t} \qquad (9)$$

By letting initial condition in Eq. (9) and multiplying by $\sin\left(\tfrac{m\pi x}{L}\right)$ then integrating from 0 to L, implies that

$$\int_0^L \sin\left(\tfrac{m\pi x}{L}\right) U(x,0) dx = \int_0^L \sum_{n=1}^{\infty} c_n \sin\left(\tfrac{n\pi x}{L}\right) \sin\left(\tfrac{m\pi x}{L}\right) \qquad (10)$$

With use of orthogonality in sin functions of Eq. (10), finally the following equation is resulted:

$$c_m = \tfrac{2}{L} \int_0^L \sin\left(\tfrac{m\pi x}{L}\right) U(x,0) dx \qquad (11)$$

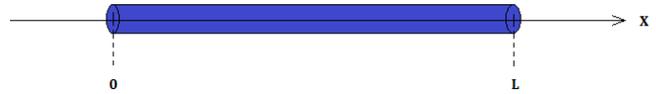

Fig -1: Coordinate system of the rod

## III. NUMERICAL SOLUTION USING FINITE DIFFERENCE METHOD

### A. Expression of Steady State Solution

In this section numerical solution of heat diffusion for the following system is presented

$$u_{xx} = \sin(x) \qquad (12)$$
$$u(x=0) = u(x=l) = 0 \qquad (13)$$

System of equations required to obtain the numerical solution. The objective of numerical solution is to determine the answer over specific certain points without having a function for it. For this purpose, the solution domain is divided into the number of intervals. The following are the procedures of the solution:

- Assume that there are N points, numbering from 0 thus there are $N-1$ intervals, $(j = 0, \ldots, N-1)$

$$\Delta X_N = \tfrac{L}{N-1} \qquad (14)$$

- Now the derivative must be determined approximately over these points, which is written as

$$\frac{d^2 U}{dX^2} \cong \frac{U_{j+1} - 2U_j + U_{j-1}}{(\Delta X_N)^2} \qquad (15)$$

- Finally substitute the mentioned approximation in differential equation

$$\frac{U_{j+1} - 2U_j + U_{j-1}}{(\Delta X_N)^2} = \sin_j \qquad (16)$$

Therefore Eq. (16) becomes

$$j = 1 \rightarrow -2U_1 + U_2 + U_0 = \sin_1 (\Delta X_1)^2 \qquad (17)$$
$$j = 2 \rightarrow U_1 - 2U_2 + U_3 = \sin_2 (\Delta X_2)^2 \qquad (18)$$

$j = N - 2 \rightarrow U_{N-1} - 2U_{N-2} + U_{N-3} = \sin_{N-1}(\Delta X_{N-1})^2$ (19)

which can be solved by using Thomas algorithm

$$\begin{bmatrix} -2 & \cdots & 0 \\ \vdots & \ddots & \vdots \\ 0 & \cdots & -2 \end{bmatrix} \begin{bmatrix} U_1 \\ \vdots \\ U_{N-1} \end{bmatrix} = \begin{bmatrix} \sin_1(\Delta X_1)^2 - U_0 \\ \vdots \\ \sin_{N-2}(\Delta X_{N-2})^2 - U_{N-1} \end{bmatrix}$$ (20)

Since a three point approximation is applied thus tri-polar system in Eq.(20) is produced. Notably, increasing the degree of accuracy results in increasement of the calculation costs.

## IV. NUMERICAL SOLUTION USING HOMOTOPY PERTURBATION METHOD

### A. Outline of the homotopy Perturbation Method Fundamentals

Assume that A and B are the neighbourhood spaces. According to definition Q is homotopic to S when there is continuous plan of

$$Q: A \times [0,1] \rightarrow B \quad \begin{cases} Q(a,0) = q(a) & ; a \in A \\ Q(a,1) = s(a) & ; a \in A \end{cases}$$ (21)

Consider the PDE and its boundary condition are expressed as

$PDE: G(u) - f(r) = 0 \quad ; r \in \Omega$ (22)
$BC: C\left(u, \frac{\partial u}{\partial \eta}\right) = 0 \quad ; r \in \Gamma$ (23)

Where $G$ is the differential factor, f is the analytical function and Γ is the boundary of PDE's domain.

When factor splits, it becomes:
$G = L + N$ (24)

Thus the Eq. (22) is written as
$L(v) + N(v) - f(r) = 0$ (25)

Now using homotopy procedure, the following expressions are written
$H(v,p): \Omega \times [0,1] \rightarrow R$ (26)

which comply
$H(v,p) = L(v) - L(u_0) + pL(u_0) + p(N(v) - f(r)) = 0$ (27)

when
$\begin{cases} p \in [0,1] \\ r \in \Omega \end{cases}$ (28)

In Eq. (27) $u_0$ defined as initial approximation of Eq.(22) and from Eq.(28) the following expression is given:
$H(v,0) = L(v) - L(u_0) = 0$ (29)
$H(v,1) = G(v) - f(r) = 0$ (30)

Substituting Eq. (29) in Eq. (25) the following equation is derived
$H(v,1) = G(v) - f(r) = 0$ (31)

Homotopic result is obtained and Eq.(27) are given by $p$ series:
$v = p^0 v_0 + p^1 v_1 + p^2 v_2 + p^3 v_3 + \cdots + p^n v_n$ (32)

Finally the approximate solution of Eq. (22) is
$u = \lim_{p \to i} v = v_0 + v_1 + \cdots + v_n$ (33)

### B. Transient State Solution Selection

The HPM solution for the system of Eq. (2) is offered with initial condition inspired from Eq. (12) and boundary conditions of Eq. (8), so the system is written as

$PDE: \frac{\partial U}{\partial t} - k \frac{\partial^2 U}{\partial x^2} = 0$ (34)

$IC: U(x,0) = \sin(x)$ (35)

$BC: U(0,t) = U(L,t) = 0$ (36)

Homotopy perturbation method procedures for Eq.(34) solution are given by:

$(1-p)\left[\frac{\partial U}{\partial t} - \frac{\partial u_0}{\partial t}\right] + p\left[\frac{\partial U}{\partial t} - k\frac{\partial^2 U}{\partial x^2}\right] = 0$ (37)

$\frac{\partial U}{\partial t} - \frac{\partial u_0}{\partial t} - p\frac{\partial u_0}{\partial t} - kp\frac{\partial^2 U}{\partial x^2} = 0$ (38)

$\frac{\partial U}{\partial t} - kp\frac{\partial^2 U}{\partial x^2} = 0$ (39)

$\frac{\partial(U_0 + pU_1 + p^2 U_2 + p^3 U_3 + \cdots)}{\partial t} - kp\frac{\partial^2(U_0 + pU_1 + p^2 U_2 + p^3 U_3 + \cdots)}{\partial x^2} = 0$ (40)

$\frac{\partial U_0}{\partial t} = 0$ (41)

Since the 0-th order of p is expressed in Eq. (41), thus
$U_0(x,t) = \sin(x)$ (42)

The 1st order of $p$ is written as:
$\frac{\partial U_1}{\partial t} - k\frac{\partial^2 U_0}{\partial x^2}$ (43)
$\frac{\partial U_1}{\partial t} + k\sin(x) = 0$ (44)
$U_1(x,t) = \sin(x) - k\sin(x)t$ (45)

And 2nd order of $p$ is expressed as
$\frac{\partial U_2}{\partial t} - k\frac{\partial^2 U_1}{\partial x^2} = 0$ (46)
$U_2(x,t) = \sin(x) - k\sin(x)t + 16\pi^4 k^2 \sin(x)t^2$ (47)

Finally,
$U(x,t) = \sin(x) e^{-kt}$ (48)

## V. RESULT AND DISCUSSION

The preceding formulations in the 4th section are used to approximate the heat diffusion phenomena in a thin rod which table (1) lists the approximate results using homotopy perturbation method. Also the exact results using analytical method for a certain values of x and t are listed in this table in order to present the relative error values.

Table -1: Results using HPM, exact values and relative errors

| X | $U_{HPM}$ ($t = 0.1$) | $U_{exact}$ ($t = 0.1$) | $\epsilon_{relative}$ |
|---|---|---|---|
| **0.25** | 0.22386035 | 0.37270783 | **0.39** |
| **0.5** | 0.43380216 | 0.52708848 | **0.17** |
| **0.75** | 0.61677225 | 0.37270783 | **0.6** |

Fig (2) presents the variation of predicted numerical solution over different values of x and t and Fig (3) shows the exact solution graph.

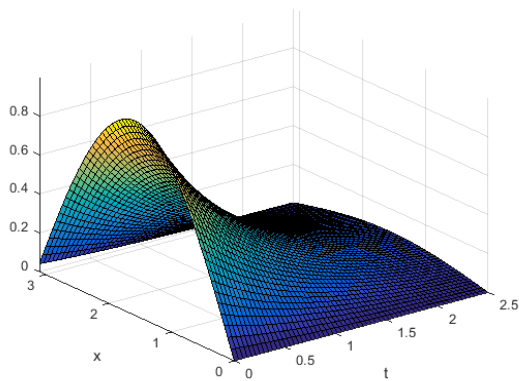

Fig -2: Variation of approximate solution over different values of x and t using HPM

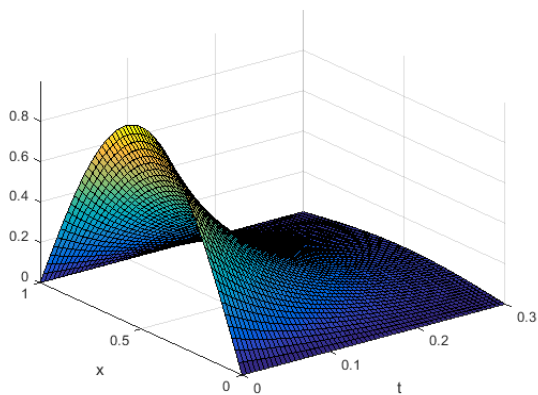

Fig -3: Exact solution over different values of x and t

The solutions in previous figures have implemented on different intervals to show the increment of absolute error close to end point when $x > {}^3\!/_4 L$ .

Table (3) lists the approximate results of one dimensional heat diffusion in a thin rod with finite difference method formulation presented in the 3$^{rd}$ section along with relative error values.

Table -3: Results using FDM, exact values and relative errors

| x | $U_{FDM}$ | $U_{exact}$ | $\epsilon_{relative}$ |
|---|---|---|---|
| 0.333 | −0.16027549 | 0.99430622 | 1.16 |
| 0.666 | −0.21370066 | 0.84049687 | 1.25 |
| 0.999 | −0.16027549 | 0.00359420 | 45.5 |

## VI. CONCLUSION

In this paper, we have successfully implemented the HPM for solving heat diffusion equation with specific boundary and initial conditions for the one dimensional thin rod shown in Fig (1). According to the results obtained, we conclude that the HPM is a powerful numerical technique for solving partial differential equations which requires less time in solution evaluation when compared to FDM.

Also it is concluded that:

- Using HPM will produce more accurate approximations to the partial differential equation solutions than FDM.

- The comparsion of these two methods reveals that the relative error in both approximations increase near end point (L), but these amounts are much fewer in HPM than FDM thus the HPM approximations are more reliable and satisfactory.

- Since HPM solutions are originated from the problems thus it is convenient to express them with different functions.